\documentclass[twocolumn,american,english]{revtex4-2}
\usepackage[T1]{fontenc}
\usepackage[latin9]{luainputenc}
\usepackage{amsmath}
\usepackage{amssymb}
\usepackage{babel}
\begin{document}
\title{Multi-radius Soler-Williams Augmented Plane Waves (SAPWMR), Multi-Radius
Soler-Williams Linearized Augmented Plane Waves (SLAPWMR) and extensions}
\author{Garry Goldstein$^{1}$}
\address{$^{1}$garrygoldsteinwinnipeg@gmail.com}
\begin{abstract}
In this work we present a new basis set for electronic structures
(Density Functional Theory (DFT)) calculations. This basis set extends
Soler Williams Linearized Augmented Plane Wave (SLAPW) basis sets
by allowing variable Muffin Tin (MT) sphere radii for the different
angular momentum channels and for different magnitude wave vectors
of the augmented plane waves. With the correct choice of MT radius,
this allows us to match additional derivatives of the wave function
at the MT radius without having to resort to additional higher derivative
terms as part of the wave function expansion inside the MT sphere.
This should lead to low wave vector basis set and low linearization
energy errors, arguably as good as APW basis set size and LAPW basis
linearization errors. We call these basis sets SAPWMR and SLAPWMR
depending on the number of derivative like wave functions kept inside
the MT radii. Similarly local orbital (LO and lo basis wave functions
depending on the number of derivative terms in the MT radius) are
suggested with a variable radius. that reduces the number of derivative
like terms needed to make them continuous or continuously differentiable.
As such semi-core states can also be well handled by our methods.
Furthermore in the appendix Full Potential Hamiltonian calculations
(FLAPW) are extended to FSLAPWMR full potential calculations for Hamiltonian
matrix elements. In the Appendix we introduce some further ideas to
improve the speed of DFT calculations which are relevant to the basis
sets presented here and to other basis sets such as the LAPW basis
set.
\end{abstract}
\maketitle

\section{\protect\label{sec:Introduction}Introduction}

In the modern environment where high throughput calculations are becoming
more and more practical and common place there is naturally significant
interest in mathematical modeling and computational methods for modern
real materials \citep{Martin_2020,Marx_2009,Singh_2006,Wills_2010,Kotliar_2006}.
Ab-initio, first principles, simulations with Density Functional Theory
(DFT) both in the Local Density Approximation (LDA) and the Generalized
Gradient Approximation (GGA) are becoming more and more commonplace
and useful \citep{Martin_2020,Singh_2006,Wills_2010,Marx_2009}. In
order to attain predictive (rather then postdictive) power for electronic
structure calculations and do systematic materials searches \citep{Martin_2020,Singh_2006}
it is imperative to increase the speed of DFT electronic structures
calculations. Many programming improvements have been made, in particular
introducing parallelizable matrix diagonalization and working on clusters
of easily as many as fifty cores or more \citep{Marx_2009,Martin_2020,Singh_2006}
that speed up calculations significantly. In this work we would like
to focus not on algorithmic improvements in computational speed but
on structural ones (ones associated with the specifics of DFT) which
are still compatible with the many programming improvements such as
parallelization.

Within these approximations (LDA and GGA), currently, there are two
main approaches to studying DFT - plane wave basis sets and pseudo
potentials and all electron calculations with augmented basis sets,
most commonly Linearized Muffin Tin Orbitals (LMTO) \citep{Skriver_1984,Andersen_1975,Andersen_1984,Andersen_2003,Hamann_1979,Wimmer_1981}
and Linearized Augmented Plane Wave (LAPW) basis sets \citep{Andersen_1975,Skriver_1984,Soler_1989,Soler_1990,Michalicek_2013,Michalicek_2014}.
Both methods strive to overcome the difficulty of finding an efficient
basis set near the atomic nucleus. Near an atomic nucleus the Khon
Sham (KS) potential is rapidly varying and requires high wave vectors
to accurately describe it, this leads to impractically large sized
secular equations for modern computers to handle efficiently. In the
pseudopotential approach the high wave vector physical KS potential
is replaced with a slowly varying one that gives the same logarithmic
derivative of the wave function at the atomic sphere radius (roughly
the radius of the atom) - that is the same boundary condition or scattering
phase shift, whereby the problem is mapped onto a more easily solvable
one. Because of the form of modern numerical Schridoinger equation
diagonalization codes, it is more efficient to use pseudopotential
methods rather then work directly with boundary conditions. In the
Augmented wave function (all electron) approach the potential is assumed
to be slowly varying in the interstitial (between the nuclei) so plane
waves (or Bessel functions multiplied by spherical harmonics) are
used there while near the nucleus (inside the MT sphere) the wavefunction
is augmented. In the Muffin Tin (MT) sphere pieces (near the nuclei)
, the exact solution to the radially averaged KS potential times a
spherical harmonic is used as the basis wavefunctions. The two pieces
of the wave function are then glued together with some boundary conditions,
usually by matching the wavefunction and some of its derivatives on
the boundary. One of the main difficulties to overcome in these all
electron methods is to find what energy to use for the solution of
the radially averaged KS potential. In the Augmented Plane Wave (APW)
methods of Slater \citep{Slater_1937} and Koringa-Khon-Rostoker (KKR)
\citep{Korringa_1947,Khon_1954} the energy is found self consistently,
which requires the solution of the radially averaged KS potential
at all energies which increases computational costs by easily an order
of magnitude \citep{Andersen_1975,Andersen_1984,Martin_2020,Marx_2009,Singh_1991,Singh_2006},
making them often impractical. In many cases it pays to linearize
the problem, say with LAPW basis sets, and focus on the solution at
some linearization energy (often in the middle of the valence band)
and study the wave function at that energy as well as its derivative
with respect to energy as a basis for the whole energy range. This
leads to manageable linearization errors and a significant speedup
of calculations. Furthermore it is possible to write down the Full
Potential Hamiltonian for this basis set, e.g. FLAPW and compute overlap
and Hamiltonian matrix elements for the KS secular self consistency
equations \citep{Wei_1985,Wei_1985(2),Wimmer_1981,Hamann_1979,Jansen_1984,Mattheis_1986,Blaha_1990}.
In some cases the valence band is not the only partially occupied
band near the Fermi energy, and as such of interest, in some cases
semi-core states which are neither completely localized nor completely
extended are essential. In this case it is advantageous to add localized
orbitals (LO and lo basis sets depending on the exact number of wave
functions kept in the MT sphere) to represent (have good overlap with)
some of these semi-core states \citep{Singh_1991,Singh_2006,Sjostdet_2000}. 

In this work we will focus on all electron LAPW like basis sets. These
are often employed as basis sets for transition metal 3d and 4d compounds
and a variety of other crystalline solids \citep{Singh_2006}. Here
we extend the Soler-Williams version of the LAPW basis set \citep{Soler_1989,Soler_1990}
to include the possibility that the MT radius depends on the angular
momentum channel and the magnitude of the plane wave wave vector considered
in the interstitial. This allows us to match an additional derivative
of the wave function at the MT sphere radius without having to resort
to higher derivative wave functions $\dot{u}_{l}=\frac{\partial}{\partial E}u_{l}$
or $\ddot{u}_{l}=\frac{\partial^{2}}{\partial E^{2}}u_{l}$ etc (here
$u_{l}$ is the solution to the spherically symmetrized Khon Sham
(KS) Hamiltonian). This feature (the ability to match many derivatives
so as to lower linearization error) and the ability not to use higher
derivative terms (the basis sets become very stiff in the presence
of higher derivative terms as the wave function becomes more and more
Bessel function like in the MT sphere) leads to a pleasant combination
of features of low basis set sizes and low linearization errors. We
will also show how to extend LO and lo to these situations where semi-core
states are important. In an Appendix we show how to extend FLAPW to
FSLAPWMR for these basis sets. In another Appendix we present further
ideas on how to improve computational speed for solid state DFT like
calculations for these basis sets and many other related ones in particular
the LAPW basis set by adjusting basis set size during iterative self
consistency loops.

\section{\protect\label{sec:SAPW}SAPWMR}

\subsection{\protect\label{subsec:Main-idea}Main idea}

Suppose you have a material with multiple atoms per unit cell. The
more correlated the atomic orbital of an atom the more important it
is to make the MT sphere of the atom it is attached to as big as possible.
However it is important that these MT spheres do not overlap limiting
the size of a possible MT sphere. Here we show how to overcome this
difficulty and optimize basis set size. This is done both by allowing
overlaps for correlated atoms and correlated angular momentum channels
and by lowering the number of derivative terms needed inside the MT
sphere. The key formula we will need is \citep{Loucks_1967}:
\begin{widetext}
\begin{equation}
\frac{1}{\sqrt{V}}\exp\left(i\left(\mathbf{k}+\mathbf{K}\right)\cdot\mathbf{r}\right)=\exp\left(i\left(\mathbf{k}+\mathbf{K}\right)\cdot\mathbf{r}_{\mu}\right)\frac{1}{\sqrt{V}}4\pi\sum_{l,m}Y_{lm}^{*}\left(\widehat{\mathbf{k}+\mathbf{K}}\right)Y_{lm}\left(\widehat{\mathbf{r-\mathbf{r_{\mu}}}}\right)i^{l}J_{l}\left(\left|\mathbf{k}+\mathbf{K}\right|\left|\mathbf{r}-\mathbf{r}_{\mu}\right|\right)\label{eq:Famous}
\end{equation}
Here $\mathbf{r}_{\mu}$ is the center of the $\mu$'th MT sphere
and $\mathbf{k}$ is a wave vector in the first Brilluoin zone while
$\mathbf{K}$ is a wave vector in the reciprocal lattice and $V$
is the volume of the unit cell in the direct lattice. This is a decomposition
of plane waves into Bessel functions times spherical harmonics and
shows that plane waves are just a sum of Bessel functions multiplied
by spherical harmonics with appropriate coefficients inside the MT
spheres. Consider the following wave functions \citep{Soler_1989,Soler_1990}:
\begin{align}
 & \breve{\chi}_{\mathbf{K}}^{\mathbf{k}}\left(\mathbf{r};\mathbf{r}_{\mu,1},....,\mathbf{r}_{\mu,M}\right)=\frac{1}{\sqrt{V}}\exp\left(i\left(\mathbf{k}+\mathbf{K}\right)\cdot\mathbf{r}\right)+\nonumber \\
 & +\sum_{\mu=1}^{M}\sum_{l,m}Y_{lm}\left(\widehat{\mathbf{r}-\mathbf{r}_{\mu}}\right)\left[A_{lm}^{\mu}u_{l\mu}^{E}\left(\mathbf{r}-\mathbf{r}_{\mu}\right)-\frac{1}{\sqrt{V}}4\pi i^{l}J_{l}\left(\left|\mathbf{k}+\mathbf{K}\right|\left|\mathbf{r}-\mathbf{r}_{\mu}\right|\right)Y_{lm}^{*}\left(\widehat{\mathbf{k}+\mathbf{K}}\right)\exp\left(i\left(\mathbf{k}+\mathbf{K}\right)\cdot\mathbf{r}_{\mu}\right)\right]\times\nonumber \\
 & \qquad\qquad\qquad\qquad\qquad\times\Theta\left(\left|\mathbf{r}-\mathbf{r}_{\mu}\right|-S_{\mu}^{l}\left(\left|\mathbf{k}+\mathbf{K}\right|\right)\right)\label{eq:SAPW}
\end{align}
\end{widetext}

Here we have subtracted out the part of the plane wave inside the
MT sphere in the second term of the second line of Eq. (\ref{eq:SAPW})
that is decomposed it as in Eq. (\ref{eq:Famous}) into Bessel functions
times spherical harmonics and subtracted them piece by piece from
the solutions of the exact radially symmetric KS equations. Indeed
here: 
\begin{equation}
\left[-\frac{d^{2}}{dr^{2}}+\frac{l\left(l+1\right)}{r^{2}}+\bar{V}_{KS}\left(r\right)\right]ru_{l\mu}^{E}\left(r\right)=Eru_{l\mu}^{E}\left(r\right)\label{eq:Schrodinger_equation}
\end{equation}
and $\bar{V}_{KS}\left(r\right)$ is the spherically average Khon
Sham (KS) potential. Here $M$ is the total number of atoms per unit
cell and $\Theta$ is the heavy-side function. Now the MT sphere radius
can depend on $l$ and the wave function is well defined even when
the MT spheres overlap. We also choose the coefficients $A_{lm}$
such that the wave function is continuous as $S_{\mu}^{l}$, whereby
we must have that:
\begin{align}
A_{lm}^{\mu} & =\frac{\frac{1}{\sqrt{V}}4\pi i^{l}J_{l}\left(\left|\mathbf{k}+\mathbf{K}\right|\left|\mathbf{r}-\mathbf{r}_{\mu}\right|\right)Y_{lm}^{*}\left(\widehat{\mathbf{k}+\mathbf{K}}\right)}{u_{l\mu}^{E}\left(S_{\mu}^{l}\left(\left|\mathbf{k}+\mathbf{K}\right|\right)\right)}\times\nonumber \\
 & \times\exp\left(i\left(\mathbf{k}+\mathbf{K}\right)\cdot\mathbf{r}_{\mu}\right)\label{eq:Continuity}
\end{align}
 Now we choose $S_{\mu}^{l}$ such that the wave function is also
continuously differentiable at $S_{\mu}^{l}$, so that the linearization
error goes as $\sim\left(\epsilon-E_{l}\right)^{4}$ \citep{Singh_2006,Martin_2020}
despite having APW basis set that is no terms proportional to $\dot{u}_{l\mu}^{E}$.

\subsection{\protect\label{subsec:Practical-methods-to}Practical methods to
find $S_{\mu}^{l}\left(\left|\mathbf{k}+\mathbf{K}\right|\right)$:
root finding using Newton's method}

We want to have that the discontinuity of the wave function's logarithmic
derivative which scales with:
\begin{widetext}
\begin{equation}
F\left(S_{\mu}^{l}\left(\left|\mathbf{k}+\mathbf{K}\right|\right)\right)\equiv J_{l}\left(\left|\mathbf{k}+\mathbf{K}\right|S_{\mu}^{l}\left(\left|\mathbf{k}+\mathbf{K}\right|\right)\right)\frac{d}{dr}u_{l\mu}^{E}\left(S_{\mu}^{l}\left(\left|\mathbf{k}+\mathbf{K}\right|\right)\right)-u_{l\mu}^{E}\left(S_{\mu}^{l}\left(\left|\mathbf{k}+\mathbf{K}\right|\right)\right)\frac{d}{dr}J_{l}\left(\left|\mathbf{k}+\mathbf{K}\right|S_{\mu}^{l}\left(\left|\mathbf{k}+\mathbf{K}\right|\right)\right)=0\label{eq:Root_finding}
\end{equation}
vanish. Here we have chosen logarithmic derivatives rather then usual
ones as there is a free coefficient $A_{lm}^{\mu}$ that is scaled
out by the logarithmic derivative and need not be considered for this
calculation. Now the derivative of this expression $F\left(S_{\mu}^{l}\left(\left|\mathbf{k}+\mathbf{K}\right|\right)\right)$
with respect to $S_{\mu}^{l}\left(\left|\mathbf{k}+\mathbf{K}\right|\right)$
is given by: 
\begin{equation}
\frac{\partial F\left(S_{\mu}^{l}\left(\left|\mathbf{k}+\mathbf{K}\right|\right)\right)}{\partial S_{\mu}^{l}\left(\left|\mathbf{k}+\mathbf{K}\right|\right)}=J_{l}\left(\left|\mathbf{k}+\mathbf{K}\right|S_{\mu}^{l}\left(\left|\mathbf{k}+\mathbf{K}\right|\right)\right)\frac{d^{2}}{dr^{2}}u_{l\mu}^{E}\left(S_{\mu}^{l}\left(\left|\mathbf{k}+\mathbf{K}\right|\right)\right)-\frac{d^{2}}{dr^{2}}J_{l}\left(\left|\mathbf{k}+\mathbf{K}\right|S_{\mu}^{l}\left(\left|\mathbf{k}+\mathbf{K}\right|\right)\right)u_{l\mu}^{E}\left(S_{\mu}^{l}\left(\left|\mathbf{k}+\mathbf{K}\right|\right)\right)\label{eq:Derivative}
\end{equation}
\end{widetext}

Now we apply Newtons method iteratively by writing: 
\begin{equation}
S_{\mu}^{l}\left(\left|\mathbf{k}+\mathbf{K}\right|\right)_{j+1}=S_{\mu}^{l}\left(\left|\mathbf{k}+\mathbf{K}\right|\right)_{j}-\frac{F\left(S_{\mu}^{l}\left(\left|\mathbf{k}+\mathbf{K}\right|\right)_{j}\right)}{\frac{\partial F\left(S_{\mu}^{l}\left(\left|\mathbf{k}+\mathbf{K}\right|\right)_{j}\right)}{\partial S_{\mu}^{l}\left(\left|\mathbf{k}+\mathbf{K}\right|\right)_{j}}}\label{eq:Iteration}
\end{equation}
to converge to the point where the discontinuity is zero by Newton's
root finding algorithm (here $j$ is the $j$'th solution and $j+1$
is the $j+1$'st solution and we iterate till sufficient convergence).
Now to speed things along we can subdivide the range of $\left|\mathbf{k}+\mathbf{K}\right|$
into a large number of points in order and use the solution of the
k'th point as the starting value of the k+1'st iteration of Newton's
method, thereby already starting close to an exact solution.

\subsection{\protect\label{subsec:Argument-why-the}Argument why the convergent
basis set will be small}

We notice that the LAPW method is based on the idea that we may well
expand many wave functions in the $l$'th angular momentum channel
inside the $\mu$'th MT sphere in the basis of $u_{l\mu}^{E}\left(r\right)$
and $\dot{u}_{l\mu}^{E}\left(r\right)$ and write: 
\begin{equation}
\psi_{l\mu}\left(r\right)=\left\langle u_{l\mu}^{E}\mid\psi_{l\mu}\right\rangle u_{l\mu}^{E}\left(r\right)+\left\langle \dot{u}_{l\mu}^{E}\mid\psi_{l\mu}\right\rangle \dot{u}_{l\mu}^{E}\left(r\right)+...\label{eq:Expansion}
\end{equation}

Now for practical coding purposes we do not wish to compute overlaps
but instead we note that any solution of the exact KS Hamiltonian
must be smooth everywhere including at the MT sphere radius and replace
the overlap condition with the condition the basis wave functions
are continuous and continuously differentiable on the MT boundary.
In the multi-radius case we still maintain the same expansion as in
Eq. (\ref{eq:Expansion}) except we choose the radius for our expansion
such that $\left\langle \dot{u}_{l\mu}^{E}\mid\psi_{l\mu}\right\rangle \cong0$
that is the wave function is continuous and continuously differentiable
at the MT radius $S_{\mu}^{l}\left(\left|\mathbf{k}+\mathbf{K}\right|\right)$
but has only the $u_{l\mu}^{E}\left(r\right)$ component are needed
to maintain the continuity of the wave function and its derivative.
This leads to a combination of the LAPW features of small linearization
error and the APW feature of small basis sets.

\subsection{\protect\label{subsec:APW+lo-basis}lo basis extensions}

We would like to consider the case where there are semi-core states
(that is states that are near but below the Fermi energy but not in
the valence band). One option for these states is multi-window basis
sets \citep{Singh_2006}where a second or more linearization energy
is chosen and the basis is doubled or more. A more efficient method
is the localized orbitals lo basis set \citep{Singh_1991,Singh_2006,Sjostdet_2000}
where specific angular momentum channels associated with these semi-core
states are chosen and augmented with additional degrees of freedom.
Here we can also do a variation on lo states (say lomr states) where
we choose 
\begin{equation}
\chi_{\tilde{lo}}^{lm,\mu}\left(\mathbf{r}\right)=\left\{ \begin{array}{cc}
0 & \left|\mathbf{r}-\mathbf{r}_{\mu}\right|\geq S_{\mu}^{\tilde{lo},l}\\
Y_{lm}\left(\mathbf{r}-\mathbf{r}_{\mu}\right)A_{lm}^{\mu}u_{l}^{E_{l}^{1}}\left(\mathbf{r},\mathbf{r}_{\mu}\right) & \left|\mathbf{r}-\mathbf{r}_{\mu}\right|<S_{\mu}^{\tilde{lo},l}
\end{array}\right.\label{eq:LO_twiddle}
\end{equation}
Where we choose the radius $S_{\mu}^{\tilde{lo},l}$ so that $u_{l}^{E_{l}^{1}}\left(S_{\mu}^{\tilde{lo},l}\right)=0$
so that we have very good wave functions near the core and still some
matching near the MT surface. This leads to improved conversion as
the wave function exactly solves the spherically averaged KS Hamiltonian
inside the MT sphere and does not depend on any derivative terms,
which increase basis set size.

\section{\protect\label{sec:Motivation}SLAPWMR}

\subsection{\protect\label{subsec:Setup}Setup}

The ideas associated with SLAPWMR are similar to those of SAPWMR except
we keep more derivative terms inside the MT sphere thereby lowering
linearization errors further. Now let us introduce the following wave
functions: 
\begin{widetext}
\begin{align}
 & \breve{\chi}_{\mathbf{K}}^{\mathbf{k}}\left(\mathbf{r};\mathbf{r}_{\mu,1},....,\mathbf{r}_{\mu,M}\right)=\frac{1}{\sqrt{V}}\exp\left(i\left(\mathbf{k}+\mathbf{K}\right)\cdot\mathbf{r}\right)+\nonumber \\
 & +\sum_{\mu=1}^{M}\sum_{l,m}Y_{lm}\left(\widehat{\mathbf{r}-\mathbf{r}_{\mu}}\right)\left[A_{lm}^{\mu}u_{l}^{E}\left(\mathbf{r}-\mathbf{r}_{\mu}\right)+B_{lm}^{\mu}\dot{u^{E}}_{l}\left(\mathbf{r}-\mathbf{r}_{\mu}\right)-\frac{1}{\sqrt{V}}4\pi i^{l}J_{l}\left(\left|\mathbf{k}+\mathbf{K}\right|\left|\mathbf{r}-\mathbf{r}_{\mu}\right|\right)Y_{lm}^{*}\left(\widehat{\mathbf{k}+\mathbf{K}}\right)\exp\left(i\left(\mathbf{k}+\mathbf{K}\right)\cdot\mathbf{r}_{\mu}\right)\right]\times\nonumber \\
 & \qquad\qquad\qquad\qquad\qquad\times\Theta\left(\left|\mathbf{r}-\mathbf{r}_{\mu}\right|-S_{\mu}^{l}\left(\left|\mathbf{k}+\mathbf{K}\right|\right)\right)\label{eq:SLAPW}
\end{align}
Here $M$ is the total number of atoms per unit cell and $\Theta$
is the heavy-side function. Now the MT sphere radius can depend on
$l$ and the wave function is well defined even when the MT spheres
overlap). We also demand that: 
\begin{equation}
\left(\begin{array}{c}
A_{lm}^{\mu}\\
B_{lm}^{\mu}
\end{array}\right)=\frac{4\pi i^{l}}{\left[S_{\mu}^{l}\right]^{2}\sqrt{V}}Y_{lm}^{*}\left(\widehat{\mathbf{k}+\mathbf{K}}\right)\left(\begin{array}{c}
\dot{u}_{l}\left(S_{\mu}^{l}\right)\frac{d}{dr}J_{l}\left(\left|\mathbf{k}+\mathbf{K}\right|S_{\mu}^{l}\right)-\frac{d}{\partial r}\dot{u}_{l}\left(S_{\mu}\right)J_{l}\left(\left|\mathbf{k}+\mathbf{K}\right|S_{\mu}^{l}\right)\\
\frac{d}{dr}u_{l}\left(S_{\mu}^{l}\right)J_{l}\left(\left|\mathbf{k}+\mathbf{K}\right|S_{\mu}^{l}\right)-u_{l}\left(S_{\mu}\right)\frac{d}{dr}J_{l}\left(\left|\mathbf{k}+\mathbf{K}\right|S_{\mu}^{l}\right)
\end{array}\right)\exp\left(i\left(\mathbf{k}+\mathbf{K}\right)\cdot\mathbf{r}_{\mu}\right)\label{eq:Solutions}
\end{equation}
\end{widetext}

so that the wave functions and its first derivatives are everywhere
continuous. We now also demand that the second derivative is continuous
leading to a linearization error of $\sim\left(\epsilon-E_{l}\right)^{6}$
\citep{Singh_2006,Martin_2020}. 

\subsection{\protect\label{subsec:Practical-methods-to-1}Practical methods to
find $S_{\mu}^{l}\left(\left|\mathbf{k}+\mathbf{K}\right|\right)$:
root finding using Newton's method }
\begin{widetext}
We will use Newton's method to compute the MT radius $S_{\mu}^{l}\left(\left|\mathbf{k}+\mathbf{K}\right|\right)$.
We want:

\begin{equation}
F\left(S_{\mu}^{l}\right)\equiv\left(\begin{array}{cc}
\frac{d^{2}}{dr^{2}}u_{l}\left(S_{\mu}^{l}\right) & \frac{d^{2}}{dr^{2}}\dot{u}_{l}\left(S_{\mu}^{l}\right)\end{array}\right)\left(\begin{array}{c}
\dot{u}_{l}\left(S_{\mu}^{l}\right)\frac{d}{dr}J_{l}\left(\left|\mathbf{k}+\mathbf{K}\right|S_{\mu}^{l}\right)-\frac{d}{dr}\dot{u}_{l}\left(S_{\mu}\right)J_{l}\left(\left|\mathbf{k}+\mathbf{K}\right|S_{\mu}^{l}\right)\\
\frac{d}{dr}u_{l}\left(S_{\mu}^{l}\right)J_{l}\left(\left|\mathbf{k}+\mathbf{K}\right|S_{\mu}^{l}\right)-u_{l}\left(S_{\mu}\right)\frac{d}{dr}J_{l}\left(\left|\mathbf{k}+\mathbf{K}\right|S_{\mu}^{l}\right)
\end{array}\right)-\left[S_{\mu}^{l}\right]^{2}\frac{d^{2}}{dr^{2}}J_{l}\left(\left|\mathbf{k}+\mathbf{K}\right|S_{\mu}^{l}\right)=0\label{eq:Zero-1}
\end{equation}
Now the derivative of 
\begin{align}
\frac{\partial F\left(S_{\mu}^{l}\right)}{\partial S_{\mu}^{l}} & =\left(\begin{array}{cc}
\frac{d^{3}}{dr^{3}}u_{l}\left(S_{\mu}^{l}\right) & \frac{d^{3}}{dr^{3}}\dot{u}_{l}\left(S_{\mu}^{l}\right)\end{array}\right)\left(\begin{array}{c}
\dot{u}_{l}\left(S_{\mu}^{l}\right)\frac{\partial}{\partial r}J_{l}\left(\left|\mathbf{k}+\mathbf{K}\right|S_{\mu}^{l}\right)-\frac{\partial}{\partial r}\dot{u}_{l}\left(S_{\mu}\right)J_{l}\left(\left|\mathbf{k}+\mathbf{K}\right|S_{\mu}^{l}\right)\\
\frac{\partial}{\partial r}u_{l}\left(S_{\mu}^{l}\right)J_{l}\left(\left|\mathbf{k}+\mathbf{K}\right|S_{\mu}^{l}\right)-u_{l}\left(S_{\mu}\right)\frac{\partial}{\partial r}J_{l}\left(\left|\mathbf{k}+\mathbf{K}\right|S_{\mu}^{l}\right)
\end{array}\right)\nonumber \\
 & +\left(\begin{array}{cc}
\frac{d^{2}}{dr^{2}}u_{l}\left(S_{\mu}^{l}\right) & \frac{d^{2}}{dr^{2}}\dot{u}_{l}\left(S_{\mu}^{l}\right)\end{array}\right)\left(\begin{array}{c}
\dot{u}_{l}\left(S_{\mu}^{l}\right)\frac{d^{2}}{\partial r^{2}}J_{l}\left(\left|\mathbf{k}+\mathbf{K}\right|S_{\mu}^{l}\right)-\frac{d^{2}}{dr^{2}}\dot{u}_{l}\left(S_{\mu}\right)J_{l}\left(\left|\mathbf{k}+\mathbf{K}\right|S_{\mu}^{l}\right)\\
\frac{d^{2}}{dr^{2}}u_{l}\left(S_{\mu}^{l}\right)J_{l}\left(\left|\mathbf{k}+\mathbf{K}\right|S_{\mu}^{l}\right)-u_{l}\left(S_{\mu}\right)\frac{d^{2}}{dr^{2}}J_{l}\left(\left|\mathbf{k}+\mathbf{K}\right|S_{\mu}^{l}\right)
\end{array}\right)\nonumber \\
 & -2S_{\mu}^{l}\frac{d^{2}}{dr^{2}}J_{l}\left(\left|\mathbf{k}+\mathbf{K}\right|S_{\mu}^{l}\right)-\left[S_{\mu}^{l}\right]^{2}\frac{d^{3}}{dr^{3}}J_{l}\left(\left|\mathbf{k}+\mathbf{K}\right|S_{\mu}^{l}\right)\label{eq:Derivative-1}
\end{align}
\end{widetext}

Now we apply Newtons method iteratively by writing: 
\begin{equation}
S_{\mu}^{l}\left(\left|\mathbf{k}+\mathbf{K}\right|\right)_{j+1}=S_{\mu}^{l}\left(\left|\mathbf{k}+\mathbf{K}\right|\right)_{j}-\frac{F\left(S_{\mu}^{l}\left(\left|\mathbf{k}+\mathbf{K}\right|\right)_{j}\right)}{\frac{\partial F\left(S_{\mu}^{l}\left(\left|\mathbf{k}+\mathbf{K}\right|\right)_{j}\right)}{\partial S_{\mu}^{l}\left(\left|\mathbf{k}+\mathbf{K}\right|\right)_{j}}}\label{eq:Iteration-1}
\end{equation}
Here $j$ is the $j$'th solution and $j+1$ is the $j+1$'st solution
and we iterate till sufficient convergence. Now to speed things along
we can subdivide the range of $\left|\mathbf{k}+\mathbf{K}\right|$
into a large number of points and use the solution of the k'th point
as the starting value of the k+1'st iteration of Newton's method much
like in SAPWMR.

\subsection{\protect\label{subsec:Key-advantage}Key advantage}

By an argument very similar to section \ref{subsec:Argument-why-the}
we see that these wave functions combine the advantages of Higher
Derivative Local Orbital (HDLO) linearization energies and the advantages
of LAPW sized basis sets. 

\subsection{\protect\label{subsec:LO-extensions}LO basis extensions }

The LO basis wave functions we may be simplified: 
\begin{widetext}
\begin{equation}
\chi_{LO}^{lm,\mu}\left(\mathbf{r}\right)=\left\{ \begin{array}{cc}
0 & \left|\mathbf{r}-\mathbf{r}_{\mu}\right|\geq S_{\mu}^{LO,l}\\
Y_{lm}\left(\mathbf{r}-\mathbf{r}_{\mu}\right)\left[A_{lm}^{\mu}u_{l}^{E_{l}^{1}}\left(\mathbf{r},\mathbf{r}_{\mu}\right)+B_{lm}^{\mu}\dot{u}_{l}^{E_{l}^{1}}\left(\mathbf{r},\mathbf{r}_{\mu}\right)\right] & \left|\mathbf{r}-\mathbf{r}_{\mu}\right|<S_{\mu}^{LO,l}
\end{array}\right.\label{eq:LAPW_Wavefunction-1-1}
\end{equation}
\end{widetext}

Where both the wave function and its first derivative are continuous
everywhere (we note that $S_{\mu}^{LO,l}$ is an additional parameter
needed for this to be possible).

\section{\protect\label{sec:Conclusions}Conclusions}

In this work we have modified the APW and LAPW basis wave functions
to allow for multiple radii for the MT spheres at the same location
(with the radius depending on the angular momentum channel and the
magnitude of the wave vector). This additional degree of freedom allowed
us to match more derivatives at the MT boundary without having to
add additional higher derivative wave function terms. We believe that
this will allow for LAPW basis accuracy with APW basis set size and
HDLO basis accuracy with LAPW basis set size and so forth (See appendix
\ref{sec:Multi-radius-HDLO-APW}). LO and lo extensions have also
been proposed allowing for the study of semi-core states. In the appendix
we have extended FLAPW to FSAPWMR and FSLAPWMR allowing for full potential
calculations of the Hamiltonian and overlap matrices. In the Appendix
we also give some additional methods to increase computational speed
for DFT type calculations by adjusting basis set size during iteration
loops. In the future it would be of interest to do numerical tests
of the efficiency of these basis sets.

\appendix

\section{\protect\label{sec:Multi-radius-HDLO-APW}HDLOMR: wave functions}

It is possible to play this trick one more time that is pick wave
functions of the form \citet{Michalicek_2013,Michalicek_2014}:
\begin{widetext}
\begin{align}
 & \breve{\chi}_{\mathbf{K}}^{\mathbf{k}}\left(\mathbf{r};\mathbf{r}_{\mu,1},....,\mathbf{r}_{\mu,M}\right)\nonumber \\
 & =\frac{1}{\sqrt{V}}\exp\left(i\left(\mathbf{k}+\mathbf{K}\right)\cdot\mathbf{r}\right)+\nonumber \\
 & +\sum_{\mu=1}^{M}\sum_{l,m}Y_{lm}\left(\widehat{\mathbf{r}-\mathbf{r}_{\mu}}\right)\left[A_{lm}^{\mu}u_{l}\left(\mathbf{r}-\mathbf{r}_{\mu}\right)+B_{lm}^{\mu}\dot{u}_{l}\left(\mathbf{r}-\mathbf{r}_{\mu}\right)+C_{lm}^{\mu}\ddot{u}_{l}\left(\mathbf{r}-\mathbf{r}_{\mu}\right)-\right.\nonumber \\
 & \left.-\frac{1}{\sqrt{V}}4\pi i^{l}J_{l}\left(\left|\mathbf{k}+\mathbf{K}\right|\left|\mathbf{r}-\mathbf{r}_{\mu}\right|\right)Y_{lm}^{*}\left(\widehat{\mathbf{k}+\mathbf{K}}\right)\exp\left(i\left(\mathbf{k}+\mathbf{K}\right)\cdot\mathbf{r}_{\mu}\right)\right]\times\Theta\left(\left|\mathbf{r}-\mathbf{r}_{\mu}\right|-S_{\mu}^{l}\left(\left|\mathbf{k}+\mathbf{K}\right|\right)\right)\label{eq:HDLO}
\end{align}
\end{widetext}

Then pick $S_{\mu}^{l}\left(\left|\mathbf{k}+\mathbf{K}\right|\right)$
such that the wave function, its first, second and third derivative
are continuous in the entire volume. This greatly reduces the linearization
error in the energies to $O\left(\epsilon-E_{l}\right)^{8}$ but greatly
slows down the calculation to speeds comparable to HDLO calculations.
It is also possible to invent new forms of LO based on this idea.

\subsection{\protect\label{subsec:LO-extensions-1}LO basis extensions }

The LO basis wave functions we may be simplified: 
\begin{widetext}
\begin{equation}
\chi_{LO}^{lm,\mu}\left(\mathbf{r}\right)=\left\{ \begin{array}{cc}
0 & \left|\mathbf{r}-\mathbf{r}_{\mu}\right|\geq S_{\mu}^{LO,l}\\
Y_{lm}\left(\mathbf{r}-\mathbf{r}_{\mu}\right)\left[A_{lm}^{\mu}u_{l}^{E_{l}^{1}}\left(\mathbf{r},\mathbf{r}_{\mu}\right)+B_{lm}^{\mu}\dot{u}_{l}^{E_{l}^{1}}\left(\mathbf{r},\mathbf{r}_{\mu}\right)+C_{lm}^{\mu}\dot{u}_{l}^{E_{l}^{2}}\left(\mathbf{r},\mathbf{r}_{\mu}\right)\right] & \left|\mathbf{r}-\mathbf{r}_{\mu}\right|<S_{\mu}^{LO,l}
\end{array}\right.\label{eq:LAPW_Wavefunction-1-1-1}
\end{equation}
\end{widetext}

Where both the wave function, its first and second derivative are
continuous everywhere (we note that $S_{\mu}^{LO,l}$ is an additional
parameter needed for this to be possible).

\section{\protect\label{sec:FAPW-1}FSAPWMR}

For simplicity we will assume that the various multi-radius MT spheres
do not overlap.

\subsection{\protect\label{sec:Definition-of-Hamiltonian-2}Definition of Hamiltonian}

First we will assume continuity of wave-functions and derivatives:
\begin{align}
\Psi_{I}\left(\mathbf{r}\right) & =\Psi_{II}\left(\mathbf{r}\right)\nonumber \\
\nabla\Psi_{I}\left(\mathbf{r}\right)\cdot d\mathbf{S}_{\mu} & =\nabla\Psi_{II}\left(\mathbf{r}\right)\cdot d\mathbf{S}_{\mu}\label{eq:Strong_conditions-2}
\end{align}
Here $d\mathbf{S}_{\mu}$ is the outward pointing normal of the sphere
and $I$ means inside the MT sphere while $II$ means outside it (interstitial).
In this case we have that the Hamiltonian matrix elements are given
by:
\begin{align}
 & \left\langle \Psi_{1}\right|H_{KS}\left|\Psi_{2}\right\rangle \nonumber \\
 & =\int_{MT}\Psi_{1I}^{*}\left(\mathbf{r}\right)\left[-\triangle+J+\vec{J}\cdot\vec{\sigma}-\mu\right]\Psi_{2I}\left(\mathbf{r}\right)\nonumber \\
 & +\int_{Int}\left[\Psi_{1II}^{*}\left(\mathbf{r}\right)\left[-\triangle+J+\vec{J}\cdot\vec{\sigma}-\mu\right]\Psi_{2II}\left(\mathbf{r}\right)\right]\label{eq:matrix_elements}
\end{align}
Where at stationarity $J=V_{KS}$ the Khon-Sham potential. We will
use the two symbols interchangeably as the stress calculations need
only be done once the system is converged, that is at stationarity.

\subsection{\protect\label{sec:Wavefuncions-2}Wave functions}

Recall that the SAPW functions now with non-collinear spin order are
given by Eq. (\ref{eq:SAPW}). Here the wave-functions satisfies the
usual radial Schrödinger equation: 
\begin{equation}
\left[-\frac{\partial^{2}}{\partial\rho^{2}}+\frac{l\left(l+1\right)}{\rho^{2}}+\bar{J}_{\alpha}\left(\rho\right)-\mu\right]\rho u_{l}^{\alpha}\left(\rho\right)=E\rho u_{l}^{\alpha}\left(\rho\right)\label{eq:Schrodinger-1-2}
\end{equation}
for some appropriately chosen energy $E$ and $\rho=\left|\mathbf{r}-\mathbf{r}_{\mu}\right|$
and we have suppressed for clarity the positions index $\mathbf{r}_{\mu}$.
Where 
\begin{equation}
\bar{J}_{\alpha}\left(\rho\right)=\frac{1}{4\pi}\int d\hat{\Omega}\left[J\left(\mathbf{r}_{\mu}+\rho\hat{\Omega}\right)+\vec{J}\left(\mathbf{r}_{\mu}+\rho\hat{\Omega}\right)\cdot\vec{\sigma}_{\alpha\alpha}\right]\label{eq:Speherical_average-1-2}
\end{equation}

\subsection{\protect\label{subsec:Interstitial-2}Interstitial}

We now pick $N\gg\mathcal{N_{\mu}}$, where $\mathcal{N}_{\mu}$ is
the total number of angular momentum channels which are not Bessel
functions in $\mu$'th MT sphere. Now let us pick $\mathbf{K}_{1}$
and $\mathbf{K}_{2}$, Now let us order the $2\mathcal{N_{\mu}}$
radii by at each $\mu$.
\begin{equation}
S_{\mu}^{l_{1}}\left(\left|\mathbf{k}+\mathbf{K}_{i}\right|\right)\leq S_{\mu}^{l_{2}}\left(\left|\mathbf{k}+\mathbf{K}_{j}\right|\right)\leq....\leq S_{\mu}^{l_{2\mathcal{N_{\mu}}}}\left(\left|\mathbf{k}+\mathbf{K}_{m}\right|\right)\label{eq:Radii-2}
\end{equation}
Now let us call the interstitial region by 
\begin{equation}
\cap_{\mu}\left|\mathbf{r}-\mathbf{r}_{\mu}\right|\geq S_{\mu}^{l_{2\mathcal{N}_{\mu}}}\left(\left|\mathbf{k}+\mathbf{K}_{m}\right|\right)\label{eq:Interstitial-2}
\end{equation}
Now for $l\leq N$ we have that:
\begin{equation}
u_{l}^{\alpha}\left(\mathbf{r}-\mathbf{r}_{\mu}\right)=\left\{ \begin{array}{cc}
u_{l}^{\alpha}\left(\mathbf{r}-\mathbf{r}_{\mu}\right) & \left|\mathbf{r}-\mathbf{r}_{\mu}\right|<S_{\mu}^{l}\left(\left|\mathbf{k}+\mathbf{K}\right|\right)\\
\frac{J_{l}\left(\left|\mathbf{k}+\mathbf{K}\right|\left|\mathbf{r}-\mathbf{r}_{\mu}\right|\right)}{\frac{J_{l}\left(\left|\mathbf{k}+\mathbf{K}\right|S_{\mu}^{l}\left(\left|\mathbf{k}+\mathbf{K}\right|\right)\right)}{u_{l}^{\alpha}\left(S_{\mu}^{l}\left(\left|\mathbf{k}+\mathbf{K}\right|\right)\right)}} & \left|\mathbf{r}-\mathbf{r}_{\mu}\right|\geq S_{\mu}^{l}\left(\left|\mathbf{k}+\mathbf{K}\right|\right)
\end{array}\right.\label{eq:u_function-2}
\end{equation}
Now for $l>N$ we define 
\begin{align}
u_{l}^{\alpha}\left(\mathbf{r}-\mathbf{r}_{\mu}\right) & =J_{l}\left(\left|\mathbf{k}+\mathbf{K}\right|\left|\mathbf{r}-\mathbf{r}_{\mu}\right|\right)\nonumber \\
A_{lm}^{\mu\alpha}\left(\mathbf{k}+\mathbf{K}\right) & =\frac{1}{\sqrt{V}}4\pi i^{l}Y_{lm}^{*}\left(\widehat{\mathbf{k}+\mathbf{K}}\right)\label{eq:Bessel_function-2}
\end{align}

\subsubsection{\protect\label{subsec:Main-setup-2}Main setup}

We write: 
\begin{align}
J\left(\mathbf{r}\right) & =\left\{ \begin{array}{cc}
\sum_{\mathbf{K}}J_{I}^{\mathbf{K}}\exp\left(i\mathbf{K}\cdot\mathbf{r}\right) & Interstitial\\
\sum_{l,m}J_{l,m}\left(\left|\mathbf{r}\right|\right)Y_{l,m}\left(\hat{\mathbf{r}}\right) & MT
\end{array}\right.\label{eq:V_KS-2}\\
\vec{J}\left(\mathbf{r}\right) & =\left\{ \begin{array}{cc}
\sum_{\mathbf{K}}\vec{J}_{I}^{\mathbf{K}}\exp\left(i\mathbf{K}\cdot\mathbf{r}\right) & Interstitial\\
\sum_{l,m}\vec{J}_{l,m}\left(\left|\mathbf{r}\right|\right)Y_{l,m}\left(\hat{\mathbf{r}}\right) & MT
\end{array}\right.
\end{align}
an similarly for the vectorized quantities. Then we have that:
\begin{widetext}
\begin{align}
\left[\bar{O}_{\mathbf{k}}\right]_{\mathbf{K},\mathbf{K}'}^{\alpha\beta} & =\left[\bar{O}_{\mathbf{k}}\right]_{\mathbf{K},\mathbf{K}'}^{I\alpha\beta}+\left[\bar{O}_{\mathbf{k}}\right]_{\mathbf{K},\mathbf{K}'}^{MT\alpha\beta}\nonumber \\
\left[-\bar{\triangle}^{\mathbf{k}}+\bar{J}^{\mathbf{k}}+\bar{\vec{J}}^{\mathbf{k}}\cdot\vec{\sigma}\right]_{\mathbf{K},\mathbf{K'}}^{\alpha\beta} & =\left[-\bar{\triangle}^{\mathbf{k}}+\bar{J}^{\mathbf{k}}+\bar{\vec{J}}^{\mathbf{k}}\cdot\vec{\sigma}\right]_{\mathbf{K},\mathbf{K'}}^{I\alpha\beta}+\left[-\bar{\triangle}^{\mathbf{k}}+\bar{J}^{\mathbf{k}}+\bar{\vec{J}}^{\mathbf{k}}\cdot\vec{\sigma}\right]_{\mathbf{K},\mathbf{K'}}^{MT\alpha\beta}\label{eq:MT_Interstitial-1}
\end{align}
Now in the interstitial we have that
\begin{align}
\left[\bar{O}_{\mathbf{k}}\right]_{\mathbf{K},\mathbf{K}'}^{I\alpha\beta} & =\Theta_{\mathbf{K}-\mathbf{K}'}\delta_{\alpha\beta}\nonumber \\
\left[-\bar{\triangle}^{\mathbf{k}}+\bar{J}^{\mathbf{k}}+\bar{\vec{J}}^{\mathbf{k}}\cdot\vec{\sigma}\right]_{\mathbf{K},\mathbf{K'}}^{I\alpha\beta} & =\left[\left(J_{KS}+\vec{J}_{KS}\cdot\vec{\sigma}\right)^{\alpha\beta}\Theta\right]_{\mathbf{K}-\mathbf{K}'}+\frac{1}{2m}\left[\mathbf{k}+\mathbf{K}'\right]^{2}\Theta_{\mathbf{K}-\mathbf{K}'}\delta_{\alpha\beta}\label{eq:Hamiltonian_interstitial}
\end{align}
Where 
\begin{equation}
\left[F\Theta\right]_{\mathbf{K}}=\sum_{\mathbf{K}'}\left[F\right]_{\mathbf{K}'}\Theta_{\mathbf{K}-\mathbf{K}'}\label{eq:convolutions-2}
\end{equation}
and 
\begin{equation}
\Theta_{\mathbf{K}}=\left(\delta_{\mathbf{K},0}-\sum_{\mu}\exp\left(-i\mathbf{K}\cdot\mathbf{r}_{\mu}\right)\frac{\left(4\pi S_{\mu}^{l_{2\mathcal{N_{\mu}}}}\left(\left|\mathbf{k}+\mathbf{K}_{m}\right|\right)\right)^{3}}{V}\cdot\frac{j_{1}\left(\left|\mathbf{K}\right|S_{\mu}^{l_{2\mathcal{N_{\mu}}}}\left(\left|\mathbf{k}+\mathbf{K}_{m}\right|\right)\right)}{\left|\mathbf{K}\right|S_{\mu}^{l_{2\mathcal{N_{\mu}}}}\left(\left|\mathbf{k}+\mathbf{K}_{m}\right|\right)}\right)\label{eq:Theta_function-2}
\end{equation}
\end{widetext}

\subsection{\protect\label{subsec:MT-pieces-1}MT pieces}

We write:
\begin{align}
 & \left[-\bar{\triangle}^{\mathbf{k}}+\bar{J}^{\mathbf{k}}+\bar{\vec{J}}^{\mathbf{k}}\cdot\vec{\sigma}\right]_{\mathbf{K},\mathbf{K'}}^{MT\alpha\beta}\nonumber \\
 & =\sum_{\mu}\sum_{l,m}\sum_{l'm'}A_{lm}^{\alpha}\left(\mathbf{k}+\mathbf{K}\right)^{*}\cdot t_{lm,l'm'}^{\mu,\phi,\phi\alpha\beta}\cdot A_{l'm'}^{\beta}\left(\mathbf{k}+\mathbf{K}'\right)\label{eq:KS_MT}
\end{align}
Where there no sum over repeated indices and 
\begin{equation}
t_{lm,l'm'}^{\mu,\phi,\phi\alpha\beta}=\sum_{l"}I_{l'll"}^{\mu\alpha\beta}G_{l'll"}^{m'mm"}\label{eq:t_matrix-1}
\end{equation}
and:
\begin{align}
G_{l,l',l"}^{m,m',m"} & =\int Y_{l,m}^{*}Y_{l',m'}Y_{l",m"}d\Omega,\nonumber \\
I_{l,l',l"}^{\mu\alpha\beta} & =\int u_{l\alpha}^{*}\left(r\right)\left[J_{l"}^{\mu}\left(r\right)+\vec{J}_{l"}^{\mu}\cdot\vec{\sigma}\right]u_{l'\beta}\left(r\right)r^{2}dr\label{eq:Definitions-1}
\end{align}

\section{\protect\label{sec:FSLAPW-1}FSLAPWMR}

For simplicity we will assume that the various multi-radius MT spheres
do not overlap.

\subsection{\protect\label{sec:Definition-of-Hamiltonian-1-1}Definition of Hamiltonian}

This is similar to Section \ref{sec:Definition-of-Hamiltonian-2}
and will not be repeated.

\subsection{\protect\label{sec:Wavefuncions-1-1}Wave functions}

This is similar to Section \ref{sec:Wavefuncions-2} and will not
be repeated.

\subsection{\protect\label{subsec:Interstitial-1-1}Interstitial}

We note that $N\gg\mathcal{N_{\mu}}$. Now let us pick $\mathbf{K}_{1}$
and $\mathbf{K}_{2}$, Now let us order the $2\mathcal{N_{\mu}}$
radii by at each $\mu$.
\begin{equation}
S_{\mu}^{l_{1}}\left(\left|\mathbf{k}+\mathbf{K}_{i}\right|\right)\leq S_{\mu}^{l_{2}}\left(\left|\mathbf{k}+\mathbf{K}_{j}\right|\right)\leq....\leq S_{\mu}^{l_{2\mathcal{N_{\mu}}}}\left(\left|\mathbf{k}+\mathbf{K}_{m}\right|\right)\label{eq:Radii-1-1}
\end{equation}
Now let us call the interstitial region by 
\begin{equation}
\cap_{\mu}\left|\mathbf{r}-\mathbf{r}_{\mu}\right|\geq S_{\mu}^{l_{2\mathcal{N}_{\mu}}}\left(\left|\mathbf{k}+\mathbf{K}_{m}\right|\right)\label{eq:Interstitial-1-1}
\end{equation}
Now for $l\leq N$ we have that:
\begin{equation}
u_{l}^{\alpha}\left(\mathbf{r}-\mathbf{r}_{\mu}\right)=\left\{ \begin{array}{cc}
u_{l}^{\alpha}\left(\mathbf{r}-\mathbf{r}_{\mu}\right) & \left|\mathbf{r}-\mathbf{r}_{\mu}\right|<S_{\mu}^{l}\left(\left|\mathbf{k}+\mathbf{K}\right|\right)\\
\frac{J_{l}\left(\left|\mathbf{k}+\mathbf{K}\right|\left|\mathbf{r}-\mathbf{r}_{\mu}\right|\right)}{\frac{J_{l}\left(\left|\mathbf{k}+\mathbf{K}\right|S_{\mu}^{l}\left(\left|\mathbf{k}+\mathbf{K}\right|\right)\right)}{u_{l}^{\alpha}\left(S_{\mu}^{l}\left(\left|\mathbf{k}+\mathbf{K}\right|\right)\right)}} & \left|\mathbf{r}-\mathbf{r}_{\mu}\right|\geq S_{\mu}^{l}\left(\left|\mathbf{k}+\mathbf{K}\right|\right)
\end{array}\right.\label{eq:u_function-1-2}
\end{equation}
Where furthermore 
\begin{equation}
\dot{u}_{l}^{\alpha}\left(\mathbf{r}-\mathbf{r}_{\mu}\right)=\left\{ \begin{array}{cc}
\dot{u}_{l}^{\alpha}\left(\mathbf{r}-\mathbf{r}_{\mu}\right) & \left|\mathbf{r}-\mathbf{r}_{\mu}\right|<S_{\mu}^{l}\left(\left|\mathbf{k}+\mathbf{K}\right|\right)\\
0 & \left|\mathbf{r}-\mathbf{r}_{\mu}\right|\geq S_{\mu}^{l}\left(\left|\mathbf{k}+\mathbf{K}\right|\right)
\end{array}\right.\label{eq:u_function-1-1-1}
\end{equation}
Now for $l>N$ we define
\begin{align}
u_{l}^{\alpha}\left(\mathbf{r}-\mathbf{r}_{\mu}\right) & =J_{l}\left(\left|\mathbf{k}+\mathbf{K}\right|\left|\mathbf{r}-\mathbf{r}_{\mu}\right|\right)\nonumber \\
A_{lm}^{\mu\alpha}\left(\mathbf{k}+\mathbf{K}\right) & =\frac{1}{\sqrt{V}}4\pi i^{l}Y_{lm}^{*}\left(\widehat{\mathbf{k}+\mathbf{K}}\right)\nonumber \\
B_{lm}^{\mu\alpha}\left(\mathbf{k}+\mathbf{K}\right) & =0\label{eq:Def_coefficient}
\end{align}

\subsubsection{\protect\label{subsec:Main-setup-1-1}Main setup}

This is identical to Section \ref{subsec:Main-setup-2} and will not
be repeated.

\subsection{\protect\label{subsec:MT-piece-1}MT piece}
\begin{widetext}
We write: 
\begin{equation}
\left[\bar{O}_{\mathbf{k}}\right]_{\mathbf{K},\mathbf{K}'}^{MT\alpha\beta}=\sum_{\mu}\sum_{l,m}\left[A_{lm}^{\alpha}\left(\mathbf{k}+\mathbf{K}\right)^{*}A_{lm}^{\beta}\left(\mathbf{k}+\mathbf{K}'\right)\delta_{\alpha\beta}+B_{lm}^{\alpha}\left(\mathbf{k}+\mathbf{K}\right)^{*}B_{lm}^{\beta}\left(\mathbf{k}+\mathbf{K}'\right)\left\langle \dot{u}_{l}^{\alpha}\mid\dot{u}_{l}^{\beta}\right\rangle \right]\label{eq:Overlaps-2}
\end{equation}
And:
\begin{align}
\left[-\bar{\triangle}^{\mathbf{k}}+\bar{J}^{\mathbf{k}}+\bar{\vec{J}}^{\mathbf{k}}\cdot\vec{\sigma}\right]_{\mathbf{K},\mathbf{K'}}^{MT\alpha\beta} & =\sum_{\mu}\sum_{l,m}\sum_{l'm'}\left[A_{lm}^{\alpha}\left(\mathbf{k}+\mathbf{K}\right)^{*}\cdot t_{lm,l'm'}^{\mu,\phi,\phi\alpha\beta}\cdot A_{l'm'}^{\beta}\left(\mathbf{k}+\mathbf{K}'\right)+B_{lm}^{\alpha}\left(\mathbf{k}+\mathbf{K}\right)^{*}\cdot t_{lm,l'm'}^{\mu,\dot{\phi},\dot{\phi}\alpha\beta}\cdot B_{l'm'}^{\beta}\left(\mathbf{k}+\mathbf{K}'\right)+\right.\nonumber \\
 & \left.A_{lm}^{\alpha}\left(\mathbf{k}+\mathbf{K}\right)^{*}\cdot t_{lm,l'm'}^{\mu,\phi,\dot{\phi}\alpha\beta}\cdot B_{l'm'}^{\beta}\left(\mathbf{k}+\mathbf{K}'\right)+B_{lm}^{\alpha}\left(\mathbf{k}+\mathbf{K}\right)^{*}\cdot t_{lm,l'm'}^{\mu,\dot{\phi},\phi\alpha\beta}\cdot A\beta_{l'm'}\left(\mathbf{k}+\mathbf{K}'\right)\right]\label{eq:KS_MT-1-1}
\end{align}
\end{widetext}

Where there no sum over repeated indices and 
\begin{align}
t_{lm,l'm'}^{\mu,\phi,\phi\alpha\beta} & =\sum_{l"}I_{l'll"}^{\mu\alpha\beta}G_{l'll"}^{m'mm"}\nonumber \\
t_{lm,l'm'}^{\mu,\phi,\dot{\phi}\alpha\beta} & =\sum_{l"}J_{l'll"}^{\mu\alpha\beta}G_{l'll"}^{m'mm"}\nonumber \\
t_{lm,l'm'}^{\mu,\dot{\phi},\phi\alpha\beta} & =\sum_{l"}K_{l'll"}^{\mu\alpha\beta}G_{l'll"}^{m'mm"}\nonumber \\
t_{lm,l'm'}^{\mu,\dot{\phi},\dot{\phi}\alpha\beta} & =\sum_{l"}L_{l'll"}^{\mu\alpha\beta}G_{l'll"}^{m'mm"}\label{eq:t_matrices}
\end{align}
\begin{align}
G_{l,l',l"}^{m,m',m"} & =\int Y_{l,m}^{*}Y_{l',m'}Y_{l",m"}d\Omega,\nonumber \\
I_{l,l',l"}^{\mu\alpha\beta} & =\int u_{l\alpha}^{*}\left(r\right)\left[J_{l"}^{\mu}\left(r\right)+\vec{J}_{l"}^{\mu}\cdot\vec{\sigma}\right]u_{l'\beta}\left(r\right)r^{2}dr\nonumber \\
J_{l,l',l"}^{\mu\alpha\beta} & =\int u_{l\alpha}^{*}\left(r\right)\left[J_{l"}^{\mu}\left(r\right)+\vec{J}_{l"}^{\mu}\cdot\vec{\sigma}\right]\dot{u}_{l'\beta}\left(r\right)r^{2}dr\nonumber \\
K_{l,l',l"}^{\mu\alpha\beta} & =\int\dot{u}_{l}^{*}\left(r\right)\left[J_{l"}^{\mu}\left(r\right)+\vec{J}_{l"}^{\mu}\cdot\vec{\sigma}\right]u_{l'}\left(r\right)r^{2}dr\nonumber \\
L_{l,l',l"}^{\mu\alpha\beta} & =\int\dot{u}_{l}^{*}\left(r\right)\left[J_{l"}^{\mu}\left(r\right)+\vec{J}_{l"}^{\mu}\cdot\vec{\sigma}\right]\dot{u}_{l'}\left(r\right)r^{2}dr\label{eq:Gaunt_coefficients}
\end{align}

\selectlanguage{american}%

\section{Lagrange Function, basis set and running cutoff (additional ideas
to improve speed)}

The results presented in this appendix are relevant to almost any
basis set, so we will phrase it with great generality. For a practical
calculation it is important to introduce a finite basis $\left|\lambda\right\rangle $
to compute $-\frac{1}{\beta}\ln\left[Tr\left[\exp\left(-\beta H_{KS}\right)\right]\right]$
- the Helmholtz free energy of the KS Hamiltonian. We now introduce
the overlap and KS Hamiltonian matrices: 
\begin{align}
O^{\lambda\lambda'} & =\left\langle \lambda\mid\lambda'\right\rangle \nonumber \\
H_{KS}^{\lambda\lambda'} & =\int d^{d}\mathbf{r}\lambda^{*}\left(\mathbf{r}\right)\left[-\frac{\nabla^{2}}{2m}+J\left(\mathbf{r}\right)-\mu\right]\lambda'\left(\mathbf{r}\right)\label{eq:Main_matrices}
\end{align}
We now introduce the secular equation 
\begin{equation}
\sum_{\lambda'}H_{KS}^{\lambda\lambda'}V_{n}^{\lambda'}=\varepsilon_{n}\sum_{\lambda'}O^{\lambda\lambda'}V_{n}^{\lambda'}\label{eq:Secular}
\end{equation}
In which case we have that 
\begin{widetext}
\begin{align}
F_{LDA}\left(D\left(\mathbf{r}\right),J\left(\mathbf{r}\right),\mu\right) & =-\frac{1}{\beta}\sum_{n}\frac{1}{1+\exp\left(-\beta\varepsilon_{n}\right)}+\frac{e^{2}}{2}\int d^{d}\mathbf{r}d^{d}\mathbf{r}'\frac{D\left(\mathbf{r}\right)D\left(\mathbf{r}'\right)}{\left|\mathbf{r}-\mathbf{r}'\right|}+\int d^{d}\mathbf{r}D\left(\mathbf{r}\right)\varepsilon_{LDA}^{\beta}\left(D\left(\mathbf{r}\right)\right)\nonumber \\
 & -\int d^{d}\mathbf{r}J\left(\mathbf{r}\right)D\left(\mathbf{r}\right)+\mu N\label{eq:LDA_Helmholtz_secular}
\end{align}
\end{widetext}

The Helmholtz free energy Lagrange function is then extremized with
respect to $D\left(\mathbf{r}\right)$, $J\left(\mathbf{r}\right)$
and $\mu$ by an iterative solution of the self consistency equations
till convergence. Here $D\left(\mathbf{r}\right)$ is the electron
density and $J\left(\mathbf{r}\right)$ and $\mu$ are Lagrange multipliers.
For simplicity we have focused on LDA (which is not essential). 

For a practical calculation for 3D solids it is important to introduce
a basis set often indexed by $\mathbf{k}+\mathbf{K}$ with $\mathbf{k}$
in the first Brillouin zone and $\mathbf{K}$ in the reciprocal lattice
and for practical reasons is limited to $\mathbf{K}<\mathbf{K}_{max}$.
In many DFT calculations the dominant contribution to computational
time is solving Eq. (\ref{eq:Secular}) for a total of $\mathcal{N}$
times till convergence with each run taking $\sim\mathbf{K}_{max}^{9}$
time for a total of $\sim\mathcal{N}\mathbf{K}_{max}^{9}$ computations.
At each step from numerical simulations it is known that the error
from an exact solution of the extermination of the Helmholtz Lagrange
function is given by \citep{Singh_2006}: 
\begin{equation}
\Delta\varepsilon\sim\varepsilon_{0}\left[\exp\left(-A\mathfrak{n}^{\alpha}\right)+\exp\left(-B\mathbf{K}_{max}^{\zeta}\right)\right]\label{eq:Extremization_error}
\end{equation}
where we are in the $\mathfrak{n}$'th iteration. Here we propose
to introduce a $\mathbf{K}_{max}\left(\mathfrak{n}\right)$ where
\begin{equation}
B\mathbf{K}_{max}^{\zeta}\left(\mathfrak{n}\right)=A\mathfrak{n}^{\alpha}\Rightarrow\mathbf{K}_{max}\left(\mathfrak{n}\right)=\left(\frac{A}{B}\right)^{1/\zeta}\mathfrak{n}^{\alpha/\zeta}\label{eq:Running_cutoff}
\end{equation}
As such computational time will then scale as 
\begin{equation}
T\sim\int_{-\infty}^{0}dt\mathbf{K}_{max}^{9}\exp\left(9\frac{\alpha}{\zeta}t\right)\Rightarrow\mathbf{K}_{max}^{9}\left[\frac{\zeta}{9\alpha}+1\right]\label{eq:Time}
\end{equation}
Where we have estimated the last iteration loop more carefully as
it is often dominant. This leads to a computational improvement of
the scale of $\sim\frac{\mathcal{N}}{1+\frac{\zeta}{9\alpha}}$ on
top of those described in the main text.
\selectlanguage{english}%

\end{document}